# Textured organic ferroelectric films from physical vapor deposition and amorphous-to-crystalline transition

Yifan Yuan, Yuanyuan Ni, Xuanyuan Jiang, Yu Yun, Xiaoshan Xu

## Abstract

Crystallization is a key for ferroelectricity which is a collective behavior of microscopic electric dipoles. On the other hand, uncontrolled crystallization leads to uneven morphology and random crystal orientations, which undermines the application potential of ferroelectric thin films. In this work, we introduce a film fabrication method of low-temperature physical vapor deposition followed by restrained crystallization, with electrical properties monitored in real-time by *in situ* measurements. This method was adopted to fabricate films of 2-methylbenzimidazole (MBI), whose molecule crystals are proton-transfer type biaxial ferroelectrics and tend to grow into a hedgehog-shaped spherulites morphology. The *in situ* measurements confirm that the crystallization, corresponding to a clear transition of physical properties, occurs dominantly during post-deposition warming. This enables the fabrication of micron-thick films in disk-shaped morphology with one polarization axis aligned along the out-of-plane direction, while the measured spontaneous polarization and coercive field are comparable to the single-crystal values. These results mark an important advancement of film growth that is expected to benefit widely the fabrication of molecular materials films whose functional properties hinge on crystallization to achieve desirable morphology and crystallinity.

# 1 Introduction

The ferroelectricity in organic materials is in increasing demand in capacitors, piezoelectric, and memory devices, due to the advantage of being light weight, flexible, and lead- and rare-metal-free which is critical for environment-friendly devices. Significant advancements have been made in organic ferroelectric polymers, such as commercially available ferroelectric polymer poly(vinylidene fluoride) (PVDF) and its derivatives[3]. Despite being able to achieve moderate remanent polarization ($P_r$) about 10 µC/cm$^2$, they tend to suffer from a high coercive electric field ($E_C$) above 500 kV/cm, which is needed to switch the polarization for the device applications.

The last decade has seen the discovery of above-room-temperature ferroelectricity in single-component organic molecular crystals, such as croconic acid (CA, $C_5H_5N_2$)[1], 2-methylbenzimidazole (MBI, $C_8H_8N_2$)[2]. Their proton-transfer origin of ferroelectricity results in the low coercive field, high remanent polarization, and persistence of ferroelectricity up to the melting temperature. For example, croconic acid has the highest recorded $P_r$ of 21 µC/cm$^2$ among organic ferroelectrics and a small $E_c$ of about 14 kV/cm owing to its O···HO hydrogen-bond network which forms a single polar axis[1]. MBI single crystal with a $P_r$ of 5.2 µC/cm$^2$ and $E_c$ of 11 kV/cm has two polar axes[2]. Polycrystal croconic acid has a large leakage current due to the adhesion of water molecules in air[3–5], which impede its development in future electronic devices. In contrast, MBI as another candidate for organic ferroelectric materials is chemically inert in air and has modest acidity[6]. Biaxial polarization opens up more flexibility for exploiting the polarization states for device applications, compared with the organics with uniaxial polarization[2].

The fabrication of MBI thin films with the desirable quasi-two-dimensional morphology and aligned polarization axis faces many challenges. Although fabrication of MBI thin films has been demonstrated by the solution shearing[6] and the evaporation from an MBI-ethanol solution[7] methods, the former method, which produces plate-like single-crystalline thin films, requires a subtle water affinity of substrates[6], while the latter method tends to form spherulites containing needle-shaped crystallite aggregates with surface roughness comparable to that of the film thickness. In principle, the physical vapor deposition (PVD) method in high vacuum offers advantages of precise thickness control and minimal contamination[8], the low sticking factor of MBI in high vacuum undermines the effectiveness, making PVD growth of MBI mostly unexplored.

Here, we fabricate $\mu m$-thick MBI films with PVD on substrates with interdigitated Pt electrode (IDE) at a cryogenic temperature (~ 230 K) to largely enhance the sticking factor, which results in quasi-two-dimensional amorphous films after deposition. The annealing during slow warming to room temperature in low vacuum leads to crystallized films with plate-like morphology, $P_r \approx 2.5$ µC/cm$^2$, and effective $E_c \approx 87.1$ kV/cm. Furthermore, the [110] crystalline direction, which is one of the polarization axes, is mostly oriented along the out-of-plane direction, particularly in thinner films. The process can be understood as a restrained spherulite growth with a low nucleation rate, in which MBI molecules nucleate into crystallites from the amorphous phase followed by the growth of the crystallites into spherulites under the restriction along the out-of-plane dimension. The growth processes, including the deposition and the post-deposition annealing, have been monitored using *in situ* measurements of remanent polarization and resistance, which confirms the amorphous-to-crystalline transition during the annealing and the small activation energy for MBI for the transition.

## 2  Results and Discussion

### 2.1  Low-temperature deposition of the amorphous phase

MBI and CA films were deposited at temperature $T_{dep}$ = 230 K and 273 K respectively on the IDE substrates, while the electrical properties were monitored *in situ* (see the "Methods" section). All $D$-$V_{app}$ relations of MBI films are consistent with the linear behavior of capacitors while those of CA are broadened by small leakage current, as shown in Fig. S1a and b, indicating no ferroelectricity during the low-temperature deposition process. The measured resistance at 200 V is 125 GΩ and 16.5 GΩ for the MBI/IDE device with 4.6 μm MBI (230 K) and CA/IDE device with 732 nm CA (273 K) respectively.

Because the origins of ferroelectricity of MBI and CA are long-range order of the hydrogen bonds in the crystalline phases[1,2], the as-deposited films are therefore amorphous with no long-range structural order which behave paraelectrically; this is expected for low-temperature deposition of MBI and CA whose sublimation temperatures in high vacuum are around 353 K and 433 K respectively. The quenching of molecular vapor onto the low-temperature substrates suppresses the motion of molecules, leaving no sufficient time and kinetic energy for crystallization.

The paraelectric behavior of the amorphous film continues as the thickness of the films increases during the low-temperature deposition. The slope of the $D$-$V_{app}$ relations for both MBI and CA increases (Fig. S1), corresponding to increasing capacitance (Fig. 1a). Multiple MBI samples with different final thicknesses were measured and the thickness dependence of capacitance distinguishes MBI from CA; the latter increases much faster. The capacitance of devices in Fig. 1a includes three parts in parallel (see Fig. 1b): $C_{glass}$ from the glass substrate, $C_{film}$ from the film, and $C_{vacuum}$ from the vacuum contribution of the upper half-plane ($C = C_{glass} + C_{film} + C_{vacuum}$). When the film thickness increases, $C_{film}$ increases while $C_{vacuum}$ decreases, leading to higher $C$ since the dielectric constant of the films is always higher than that of vacuum.

The dielectric constant of the amorphous films can be extracted by separating the contributions to the device capacitance. We analyzed the thickness dependence using the partial-capacitance model[9,10]. Essentially, the capacitance is a function of the dielectric constant of the films, the electrode length, $L$, and of two dimensionless geometric parameters: $\eta$ as the ratio between the electrodes gap and finger widths, and $r$ as the ratio between the thickness of the film and the IDE spatial period. The function $C(\varepsilon_{MBI}, L, \eta, r)$ is summarized in the Supplementary Note 4. Fitting the thickness dependence of the device capacitance in Fig. 1a, the dielectric constants are found as $\varepsilon_{MBI}$ =2.93, $\varepsilon_{CA} = 7.28$, and $\varepsilon_{sub} = 5.63$ for amorphous MBI, amorphous CA, and the glass substrates respectively. Notice that $\varepsilon_{MBI}$ is much smaller than the single-crystal MBI value 33.2 (measured at 1 Hz, 230 K)[2], indicating the effect of long-range ferroelectric order.

### 2.2  Crystallization and emergence of ferroelectricity in post-deposition annealing.

After the deposition of MBI at 230 K, the films were heated up to 305 K at a 0.2 K/min rate, annealed at 305 K, and later cooled back down to about 260 K, in a 3 Torr N$_2$ gas environment to minimize sublimation. Increasing the pressure from vacuum to 3 Torr has no effect on the hysteresis loop, and $D_{max}$ ($D$ measured at 200 V) remains the same. In addition to the $D$-$V_{app}$ relation, the remanent polarization $P_r$ and capacitance were measured using the double wave method (DWM) (see Supplementary Note 2 and Fig. S6). As the temperature increases, the $D$-$V_{app}$ relations undergo a transition from the reversible linear behavior to the typical ferroelectric hysteresis loops as shown in Fig. 1c for a 3.1 μm film. To quantify the transition, we plot in Fig. 2a the temperature dependence of $P_r$ and $D_{max}$. Similar trends of $P_r$ and $D_{max}$ in Fig. 2a indicate the process of the phase transition. Since $D_{max}$ includes the contributions from both the dielectric layers (glass substrate and vacuum) and the ferroelectric layer (MBI film), this means that the ferroelectric contribution of MBI dominates the change of $D_{max}$ during the transition.

Since ferroelectricity relies on long-range structural order in the crystalline phase, this transition of $P_r$, which occurs in a narrow range near 290 K (Fig. 2), corresponds to an amorphous-to-crystalline structural transition, or crystallization from the amorphous phase. According to the definitions of glass transition[11], the crystallization temperature $T_x$ (see Supplementary Fig. S3) for the MBI film of 3.1 μm is 288.5 K; the corresponding end temperature $T_f$ of the transition (Fig. S3) is 293.2 K. We note that the crystallization temperature $T_x$ depends on many factors, such as film thickness[12], heating rate[13], and deposition temperature[14]. Our results indicate that $T_x$ of MBI increases linearly with the film thickness (Fig. S5). It is possible that interfacial energy between the substrate and the film plays an important role. It was also found that $T_x$ increases with decreasing film thickness for some polymers when the interfacial interaction is strongly attractive[12,15].

As the temperature increases up to $T_x$, the $D$-$V_{app}$ loop opens up slightly (Fig. 1c), consistent with leaky capacitor behavior of amorphous phase, indicating the reduction of resistance with increasing temperature. During the crystallization from 288.5 K to 293.9 K, ferroelectric polarization increased rapidly (Fig. 2a), while the hysteresis loops also exhibit visible leaky behaviors at 305 K (Fig. 1c). Above 293.9 K, the remanent polarization $P_r$ increases linearly and slowly, reaching 2.60 μC/cm$^2$ at 305 K. The hysteresis loop at 305.2 K, 1 Hz has an averaged coercive voltage ($V_c$) 44.6 V and an effective $E_c \approx 87.1$ kV/cm.

After the substrate temperature reaches 305 K, the film was annealed at 305 K for 7.9 hours in a 3 Torr N$_2$ environment. Measurement after the annealing shows a small reduction of $P_r$ and $D_{max}$ (Fig. 2b), which is consistent with the small fraction of film sublimation (with a rate of 5.7 nm/h on average) indicated by the thickness monitor. Temperature dependence of $P_r$ and $D_{max}$ were also plotted in Fig. 2a during the cooling from 305 K to about 260 K at a constant rate of 0.5 K/min. Above $T_f$, $P_r$ and $D_{max}$ reduce linearly with a rate similar to that during the warming process in the same temperature range, suggesting that this reversible temperature dependence is intrinsic to ferroelectric MBI crystal. Further cooling through $T_x$, continues the near-linear reduction of $P_r$ and $D_{max}$, in contrast to the irreversible phase transition in this temperature range during the warming process. The $D$-$V_{app}$ loop also becomes more typical for ferroelectric switching due to the reduction of leakage. In most ferroelectrics, $P_r$ decreases with temperature since it is expected to vanish at the ferroelectric-to-paraelectric phase transition temperature. However, for proton-transfer type ferroelectricity that relies on long-range order of hydrogen bonds, like MBI and CA, since the elongation of hydrogen bonds dominates the thermal expansion, a near-linear relationship between strain and ferroelectricity is expected[16], which has also been observed previously in MBI[17] and CA[1].

2.3    Electrical capacitance and resistance of MBI films during the transition

As shown in Fig. 2b, the MBI film capacitance was measured by fitting the non-switching part of the $D$-$V_{app}$ relation from the DWM measurement and subtracting the substrate capacitance, which is described in Supplementary Note 2 and 3.

It is evident from the data in Fig. 2 that both the drastic decline in resistance and the sharp increase in capacitance reflect the structural transition in the films. Before the phase transition $T_x$, the dielectric constant of the amorphous film increases slightly with temperature (Fig. 2b) which is common in amorphous materials[18]. With higher kinetic energy by thermal excitation, the bound charges in amorphous films are easier to follow the change of the external electric field. During the phase transition, the capacitance increases drastically with temperature due to the much larger dielectric constant of crystalline MBI than that of amorphous MBI. In contrast to crystalline MBI with the long-range order, the polarizability of protons is greatly reduced due to the disordered positions of molecules in amorphous states. After the completion of the phase transition, $C_{MBI}$ decreases as $T$ increases, which is unexpected. To see whether this is an artifact from the measuring method, we also measured the capacitance directly using an impedance analyzer by repeating the same thermal process on another film (1.55 μm thickness). The drop of capacitance immediately after $T_f$ also appeared but only at low frequency (Fig. S10). More detailed studies are needed to elucidate the exact mechanism.

The device resistance ($R_{device}$) at 200 V and substrate resistance were obtained from $I$-$V_{app}$ measurements *in situ* with a sweeping rate ≈0.55 V/s (≈ 0.7 mHz), in the same annealing process where the $D$-$V_{app}$ relations were measured. The film resistance ($R_{MBI}$) and the substrate resistance ($R_{sub}$) are in parallel, so $R_{MBI}$ were calculated from $R_{device}$ and $R_{sub}$ which was measured before the deposition of films (see Fig. S2). As shown in Fig. 1d, the $I$-$V$ curves at low temperature (251.1 K) before crystallization exhibit linear ohmic behaviors which can be described by a leakage resistance. Below 283 K, the glass substrates contribute significantly to the leakage resistance (Fig. S2), although the change of resistance with increasing temperature is dominated by the MBI (Fig. 2c), which can be attributed to the thermal activation of charge carriers in the amorphous phase. As shown in Fig. 1d, starting around 287.3 K, the polarization-switching peaks appear and become increasingly prominent with increasing temperature, while the linear background also increases with temperature, indicating the reduction of resistance at higher temperatures. The temperature range in which the current peaks are prominent is consistent with that of the amorphous-to-crystal transition. The resistance of MBI undergoes a rapid reduction from 4.43 GΩ at $T_x$ to 1.26 GΩ at $T_f$ (Fig. 2c). Above $T_f$ (293.1 K), most of the amorphous materials have transformed to crystals; the film resistance decreases slightly with increasing temperature.

### 2.4 Morphology and texture of crystalized MBI films

The morphology of the crystallized films consists of disk-like polycrystalline aggregates, as measured using a laser scanning microscope *ex situ* in the atmosphere at room temperature. As shown in Fig. 3 and Fig. S7, the plane view of the MBI film shows round-shaped outer envelopes of the aggregates. The height profile (Fig. 3a) can be described as a plateau with a spherical cap on top. The average height is consistent with the nominal thickness of the film, with a small height/diameter aspect ratio of ≈0.039 (Fig. 3e), suggesting that the quasi-two-dimensional morphology of the films are mostly kept during the crystallization from the amorphous phase.

A clear texture of the film, i.e., the radial alignment of the crystallites within the aggregates, can be seen in Fig. 3c. Basically, all the crystallites are of elongated shapes with the longer dimension pointing toward the center of the aggregates. A close-up view of individual crystallites is provided by the atomic force microscopy (AFM) measurements, as shown in Fig. 3b. The crystallites are clearly aligned along their longer dimension. The size of the shorter dimension is about 0.5 μm, as measured from the AFM image in Fig. 3b. This texture of radial alignment is related to the crystal growth mode of spherulites[19], which is often found in polymer films[20].

Starting with nucleation from the amorphous phase, crystallites grow through the mechanism of small-angle branching, in which numerous needle-like or fibrous crystals spike outward from the nuclei. The shape of the spherical cap at the top of the MBI aggregates (Fig. 3a) is also consistent with the spherulites growth mode, although the growth is restrained by the initial quasi-two-dimensional morphology of the amorphous films, according to the small aspect ratio.

The key feature of the film texture is the crystalline alignment of the crystallites. In spherulites, the radial direction typically corresponds to the crystalline direction of the fastest growth, which has been shown as the *c* axis of MBI by synchrotron X-ray diffraction measurements previously [6]. The elongated shape of MBI crystallites originates from its anisotropic pseudo-tetragonal crystal structure[2] and the large contrast of surface energy of different crystal planes. As illustrated in Fig. 4a, the MBI molecules, essentially lying flat in the a-b plane of the unit cell, form a chain-like structure via the hydrogen bonds. Within the same a-b plane, the chains are connected by van-de Waal bonds. On the other hand, between different MBI molecular planes, there are much more Van der Waals bonds, suggesting a larger surface energy of the (001) plane. The growth of crystals tends to minimize the (orientation-dependent) surface energy integrated over the entire crystal.[21] Therefore, the fastest growth rate is along the facet direction with the largest surface energy, which for MBI is the [001] direction (*c* axis).

This alignment of the [001] crystalline direction along the radial direction of the spherulites is corroborated by the optical images under the cross-polarized microscope (Fig. 4d), where the light passes a linear polarizer, gets reflected by the sample, and passes an analyzer whose axis is perpendicular to that of the polarizer, before finally being captured by the camera. As shown in Fig. 4d, the brightness of the spherulites is minimum when the crystallites are either parallel to the axis of the polarizer or to that of the analyzer, showing the "Maltese Cross".[22] MBI crystals, as hinted by the anisotropic crystal structure, are uniaxial birefringent materials with the primary optic axes along the c axis (the radial direction in Fig. 4d) and in the a-b plane. Defining α as the angle between the radial direction of the spherulites and the polarizer axis, if the radial direction corresponds to the c axis, α is also the angle between the c axis and the polarizer axis. As explained in Supplementary Note 5, the amplitude of the light after the analyzer is proportional to $\sin(2\alpha)$, which is consistent with the observation in Fig. 4d.

Besides the alignment of the *c* axes along the radial directions of the spherulites, the crystalline orientations pointing out of the film plane also appear to be aligned, as indicated by the x-ray diffraction measurements. To confirm the orientation of crystals, we investigated the properties of MBI with various thicknesses. As shown in Fig. 4b, the $\theta/2\theta$ x-ray diffraction (XRD), which measures the crystal planes parallel to the film plane, shows two dominant peaks $(110)_{tetra}$ and $(220)_{tetra}$ from MBI, indicating that the [110] direction is the dominant out-of-plane direction. The XRD peaks of each film are normalized to the peak intensity of $(110)_{tetra}$ planes. As the film thickness increases, the peak intensity of $(002)_{tetra}$ planes gradually increases, indicating that the *c* axes of some crystallites gradually deviate from the film plane in thicker films, consistent with the restricted spherulite growth. The change of peak intensity ratio of $(002)_{tetra}/(110)_{tetra}$ with the thickness is also in line with the trend of aspect ratio as shown in Fig. 4c.

As illustrated in Fig. 4a, MBI is a bipolar ferroelectric material whose polarization is along the in-plane $[1\bar{1}0]_{tetra}$ and $[1\bar{1}0]_{tetra}$ directions, which are also the directions of the molecular chains connected by the hydrogen bonds. Therefore, the two polar axes are either normal to the film surface or in the plane of the MBI films, suggesting that the films can be poled either along the in-plane or along the out-of-plane direction, which is consistent with the successful measurement of switchable polarization using the in-plane electrodes (see Fig. 1).

## 2.5 Low nucleation density for the restrained spherulite growth

The disk-like morphology and the radial texture of the MBI films observed in Fig. 3 and 4 rely on the low density of spherulites and the restriction of growth along the out-of-plane direction.

The restriction of the spherulite growth along the out-of-plane direction hinges on the formation of quasi-two-dimensional morphology, defined by the film thickness, during the low-temperature deposition before the crystallization process. Assuming the number of nucleation centers per unit volume is $\rho$ and each nucleation center grows into a spherulite, the average volume per spherulite is $1/\rho$. As illustrated in Fig. S12 and Fig. 4c, for the film of thickness $h$ = 3.1 um and area $A$ = 594085 μm$^2$, the number of nucleation center is $N$ = 261, corresponding to $1/\rho$ = 7056 μm$^3$ and $\rho^{-\frac{1}{3}} \gg h$. The growth of the spherulite is then restrained along the out-of-plane direction, with the radial directions of the spherulite lying mostly in the film plane. This leads to a spherical-cap-like morphology, with a small thickness/diameter aspect ratio, as observed in Fig. 3a. As shown in Fig. 4c, when the thickness increases, the restraint along the out-of-plane direction reduces, leading to the increasing aspect ratio. In contrast, for example, the vapor evaporation of MBI at room temperature in the atmosphere has no restriction along the out-of-plane direction, resulting in crystalline nuclei rather than amorphous molecules on the surface,[23] which produces three-dimensional spherulites.[7]

Clearly, the low nucleation density is the other key factor to the restrained spherulite growth. Our previous study[8] on the nucleation kinetics in physical vapor deposition of thin films indicates that the nucleation rate is a non-monotonic function of substrate temperature. At the low-temperature limit, the nucleation rate is low due to the low molecular diffusion; when the temperature is approaching the sublimation (or melting) temperature $T_m$, the nucleation rate is also low due to the diminishing undercooling. The maximum nucleation rate occurs at an intermediate temperature as a compromise of diffusion and undercooling.[8] For MBI films, the nucleation density and coverage rate decreases rapidly from 216 K to 226 K, while the nucleation density is similar between 226 K and 236 K (Fig. S7). Therefore, the deposition temperature 230 K in 10$^{-7}$ torr pressure for the MBI films in this study is in the high-temperature tail of the nucleation rate-temperature relation, which has the desirable small nucleation rate.

## 2.6 Kinetics of the anisothermal amorphous-to-crystalline phase transformation in MBI.

The kinetics of the amorphous-to-crystalline transformation depends on the characteristics of the nucleation and crystallite growth processes, involving two important parameters, time ($t$) and temperature ($T$). Under constant temperature, the kinetics of the transformation can be described by the Johnson-Mehl-Avrami (JMA) equation[14,24], i.e.,

$$f = 1 - \exp\left[-\left(\frac{t}{\tau}\right)^n\right], \tag{1}$$

where $f$ is the fractional degree to which transformation has taken place by time $t$, $n$ is the Avrami exponent whose value depends on the characteristics of nucleation and the dimensionality of the crystal growth, $\tau$ is the relaxation time satisfying an Arrhenius-type-dependence upon temperature[25], i.e.,

$$\tau = \tau_0 \exp\left(\frac{E}{k_B T}\right), \tag{2}$$

where $\tau_0$ is a temperature-independent pre-exponential factor, $E$ is the activation energy, $k_B$ is the Boltzmann constant. In this work, $T$ varies linearly with time, as described by

$$T = T_0 + kt \tag{3}$$

where $T_0$ is the initial temperature, $k = 0.2$ K/min.

Combining Eq. (1), Eq. (2), and Eq. (3), one has

$$f(T) = 1 - \exp\left[-\left(\frac{T - T_0}{k\tau_0 \exp\left(\frac{E}{k_B T}\right)}\right)^n\right] \tag{4}$$

Although the variation of conductivity or capacitance is often used [24,26] to measure the fraction of the crystalline phase, for ferroelectric MBI crystals, normalized remanent polarization is used here to measure the degree of phase transformation. Here we subtract the background temperature dependence of the remanent polarization of the MBI crystal (Fig. 2a) (see Supplementary Note 6), after which the normalized $f(T)$ is plotted in Fig. 5a.

The transformation behavior with the temperature is fitted using Eq. (4) assuming that the Avrami exponent ($n$) and the activation energy ($E$) are independent of temperature. The fit parameters are shown in Table 1 and the MATLAB script is in the separate supplementary file. To the best of our knowledge, these parameters for MBI haven't been reported before. The activation energy $E$ is the minimum kinetic energy that molecules must acquire to organize themselves. Comparing with the activation energy of other materials in Table 2, MBI molecules ($C_8H_8N_2$) have very small activation energy (0.21 eV) reflecting the fact it is much easier to crystalize from the amorphous state than inorganics and polymers with large molecules as building units.

The Avrami exponent $n$ reveals more details on the dimensionality of the crystallite growth. Avrami et. al developed a general equation for transformation kinetics of three-dimensional growth[27], $f = 1 - \exp\left[\eta Y_1 Y_2 Y_3 \int_0^t I(t - \tau)^3 d\tau\right]$, where $\eta$ is a shape factor, $Y_i$ is the anisotropic growth rate, $I$ is the nucleation rate. If the nucleation rate $I$ is constant, this equation leads to Eq. (1) with $n = 4$. In our experiments, the exponent $n$ for MBI is 4.2, indicating that the crystallite growth is three-dimensional and the nucleation rate may increase slightly with temperature.

Typically, initial nucleation occurs preferentially in sites with high interfacial energy or defects, and random nucleation will appear after the saturation of these preferential sites. In terms of crystallization of the amorphous phase, it has previously been suggested that the initial nucleation occurred in the surface layer in amorphous solid water films and the growth front propagated into deep layers,[28] due to the higher mobility of molecules in the surface layer than inside the film. Therefore, it is conceivable that MBI's initial nucleation presented in this study may start from the surface layer or the substrate interface shown in Fig. 5b.

## 3    Conclusion

In summary, this work demonstrates that low-temperature physical vapor deposition in the high vacuum is a viable method to fabricate polycrystalline MBI thin films. By exploiting the extreme anisotropic growth of MBI crystals, a preferred crystalline orientation along (110) plane and the corresponding spontaneous polarization direction of MBI thin films is engineered with the knowledge of nucleation and growth of crystals which are inevitable processes of phase transition. This study also demonstrates the detailed process of the amorphous-to-crystalline phase transition in organic molecular ferroelectrics. The crystallization temperature of amorphous MBI films is around 287 K. The phase transformation details extracted from the *in situ* electrical measurements including $D$-$V_{app}$ loops, remanent polarization, resistance, and capacitance provide insights into the anisothermal kinetics of phase transition in MBI which has a very small activation energy of 0.21 eV. The evolution of ferroelectricity in organics can be naturally extended to other organic ferroelectric materials, thus offering significant insights into the design of functional high-performance organic ferroelectric devices through molecule and crystal engineering. The electrical properties and multiple fabrication methods for MBI molecules clearly demonstrate that organic ferroelectrics may provide alternative sources for piezoelectric and ferroelectric devices.

## 4 Experimental Section

- Film sample fabrication and electrical measurement

The substrates with interdigitated Pt electrodes on glass were purchased from Metrohm; the electrode width and gap between electrodes are 6 and 4 μm respectively, as shown in Fig. S11b; there are 125 pairs of electrodes of 6760 μm long. The physical vapor deposition was performed in the high vacuum (< $10^{-6}$ torr) using an EvoVac system (Angstrom Engineering Inc.) equipped with a thickness monitor; the schematic of the system is shown in Fig. 6. The deposition was carried out with substrate temperature 230 K and 273 K, rate 1.8 and 1.2 nm/min, for MBI and CA respectively. The temperature of the thickness monitor was kept the same as the substrate during deposition. After the deposition finishes, the substrate is heated at a constant rate of 0.2 K/min in the high vacuum (< $10^{-6}$ torr) and in ≈3 torr $N_2$ environment for CA and MBI respectively. It is confirmed that the change of pressure for MBI has no effect on the hysteresis loops. $D$-$V_{app}$, DWM, and $I$-$V_{app}$ loops were measured with Precision RT66C Ferroelectric Tester (Radiant Technologies, Inc.). The frequency was set to 1 Hz to ensure saturated polarization. The electric displacement is equal to the measured charges divided by the whole electrode area. Before deposition, the $D$-$V_{app}$ loop for the substrate was measured to ensure no breakdown of electrodes (Fig. S4b). The capacitance during deposition was acquired by fitting the $D$-$V_{app}$ loops, and the capacitance during annealing from the fitting of leakage polarization in DWM measurements (Fig. S6).

- Morphology and Structural Characterization

X-ray diffraction pattern (XRD) was obtained with the Rigaku Smartlab Diffractometer at room temperature in the atmosphere by $\theta$-$2\theta$ scanning. The samples for XRD measurements are deposited at 230 K on IDE substrate with various thicknesses. The film thickness after deposition was also measured by the DektakXT stylus profiler (Bruker) to calibrate the thickness monitor. Atomic force microscopy (AFM) analysis was performed in the air by using a Bruker Dimension ICON SPM in peak force tapping mode. Optical microscopy (OM) images with height profiles were obtained with Keyence laser scanning microscope VK-X200K. The MBI sample with a thickness of 4.28 μm for OM measurements is deposited at 236 K on the IDE substrate; for AFM measurement the film was deposited at 253 K on the $Al_2O_3$ substrate.

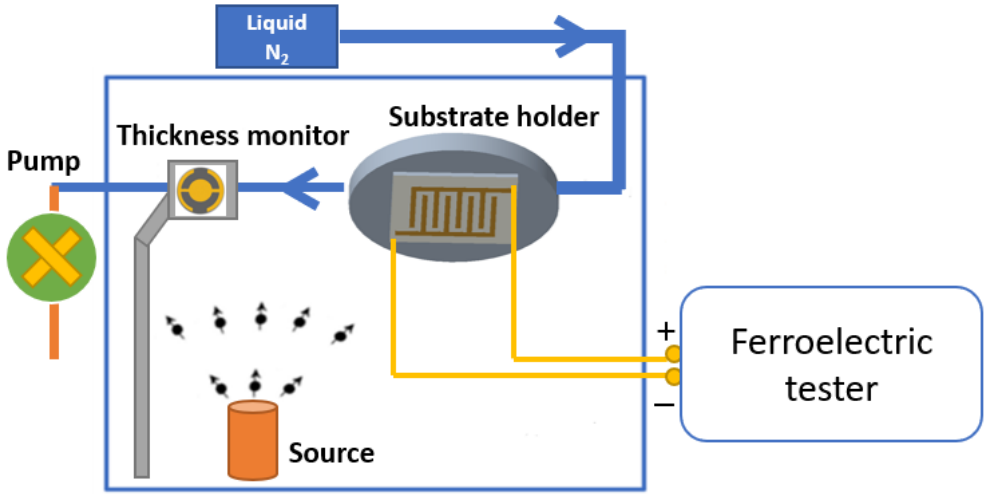

Fig. 6 Schematic of the deposition system with the cooling system..

**Acknowledgments**

This research was primarily supported by the U.S. Department of Energy (DOE), Office of Science, Basic Energy Sciences (BES), under Award No. DE-SC0019173. The research was performed in part in the Nebraska Nanoscale Facility: National Nanotechnology Coordinated Infrastructure and the Nebraska Center for Materials and Nanoscience (and/or NERCF), which are supported by the National Science Foundation under Award ECCS: 1542182, and the Nebraska Research Initiative.

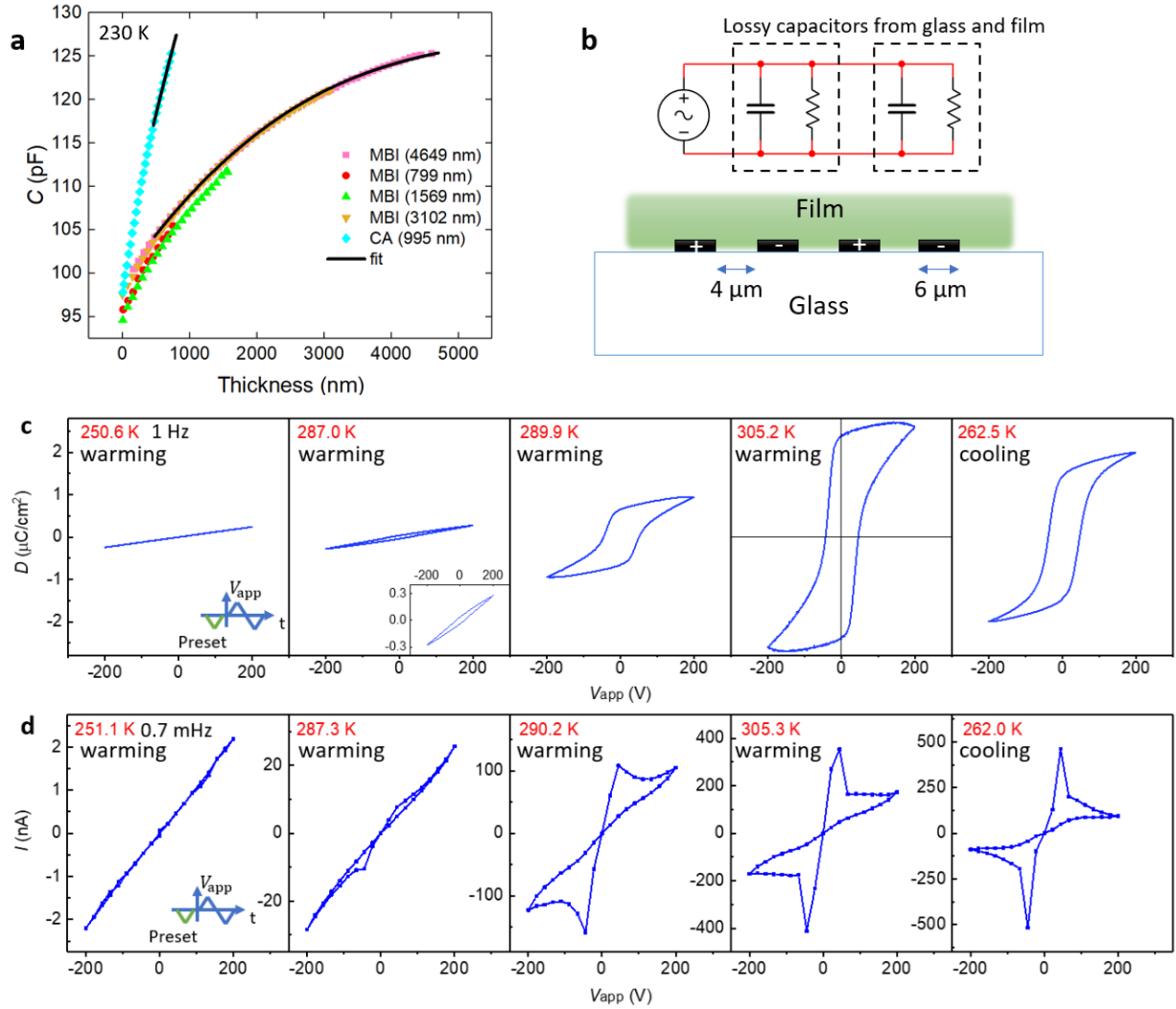

Fig. 1 a) Capacitive response of IDE devices for MBI or CA films with different final thicknesses. The fit was performed with the partial-capacitance model. Note that this model applies for $r>0.05$ ($r$ is described in Supplementary Note 4). b) Schematic of the IDE device and its equivalent circuit. c) The $D$-$V_{app}$ loops at 1 Hz, different temperatures corresponding to the red spots in Fig. 2a. The inset in (c) is the close-up view at 287 K. d) The $I$-$V_{app}$ curves at different temperatures corresponding to the red spots in Fig. 2c.

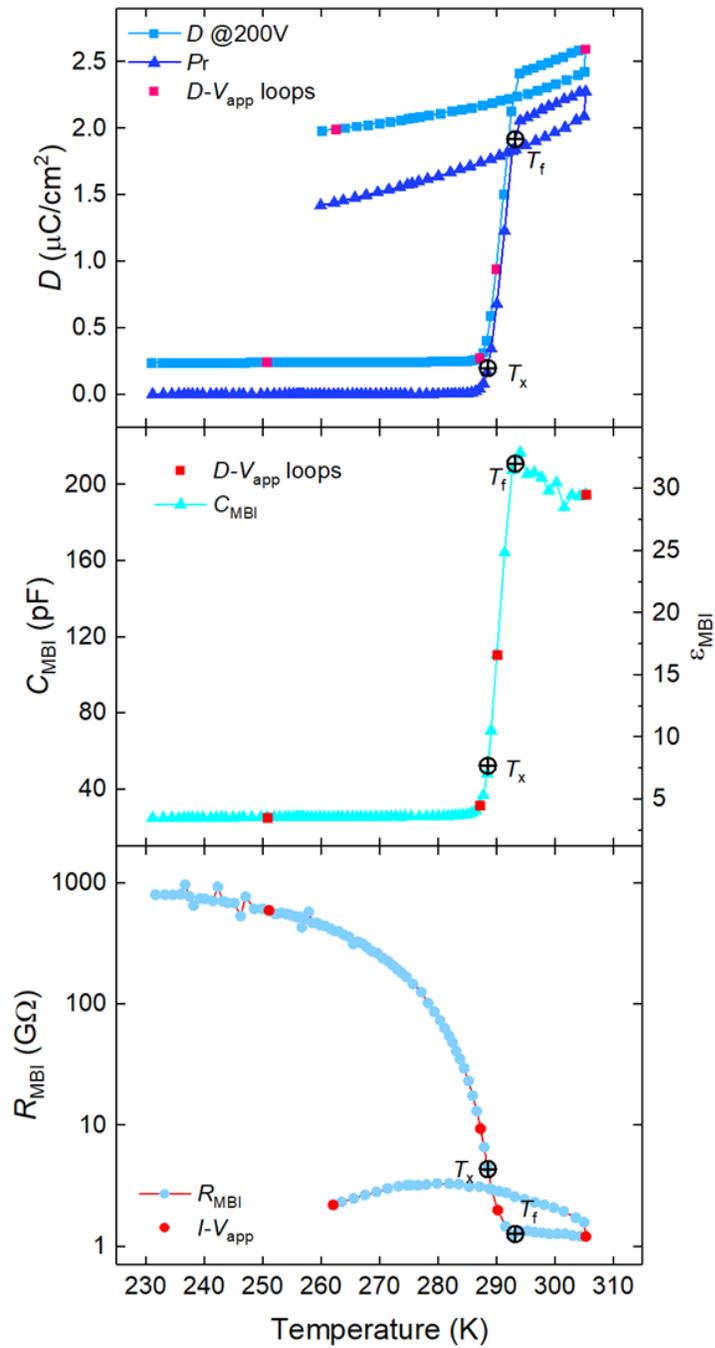

Fig. 2 Variations of the maximum electrical displacement, remanent polarization (a), capacitance (b), and resistance (c) with the temperature immediately after deposition. The thickness of the sample is 3.1 μm. The resistance is extracted from $I$-$V_{app}$ curves at 200 V.

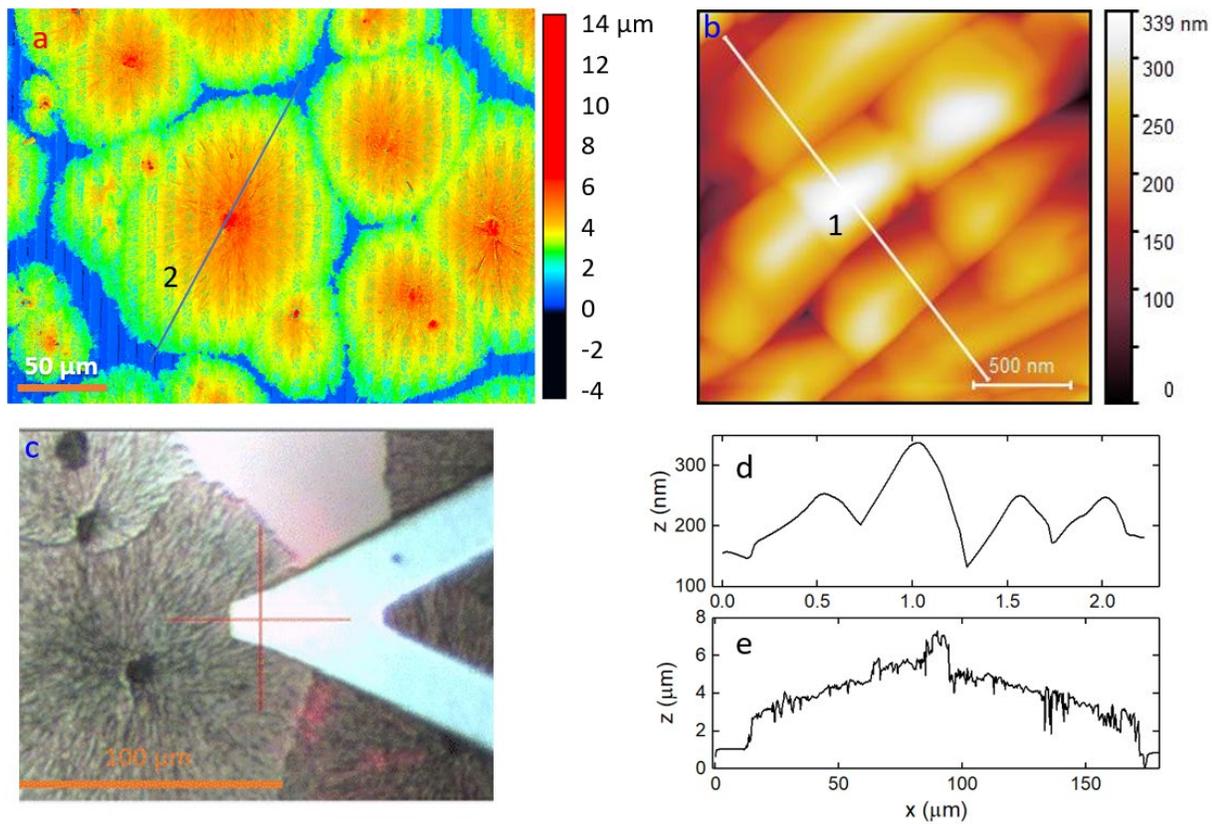

Fig. 3 Morphology of crystalline MBI films. a) Height image of the MBI film with a thickness of 4.28 μm by the laser scanning microscope. b) AFM image of the assembled spherulite of MBI at the cross position in c). In c) MBI crystals deposited on the $Al_2O_3$ substrate at 253 K. d,e) Height profiles along line 1 in b) and line 2 in a).

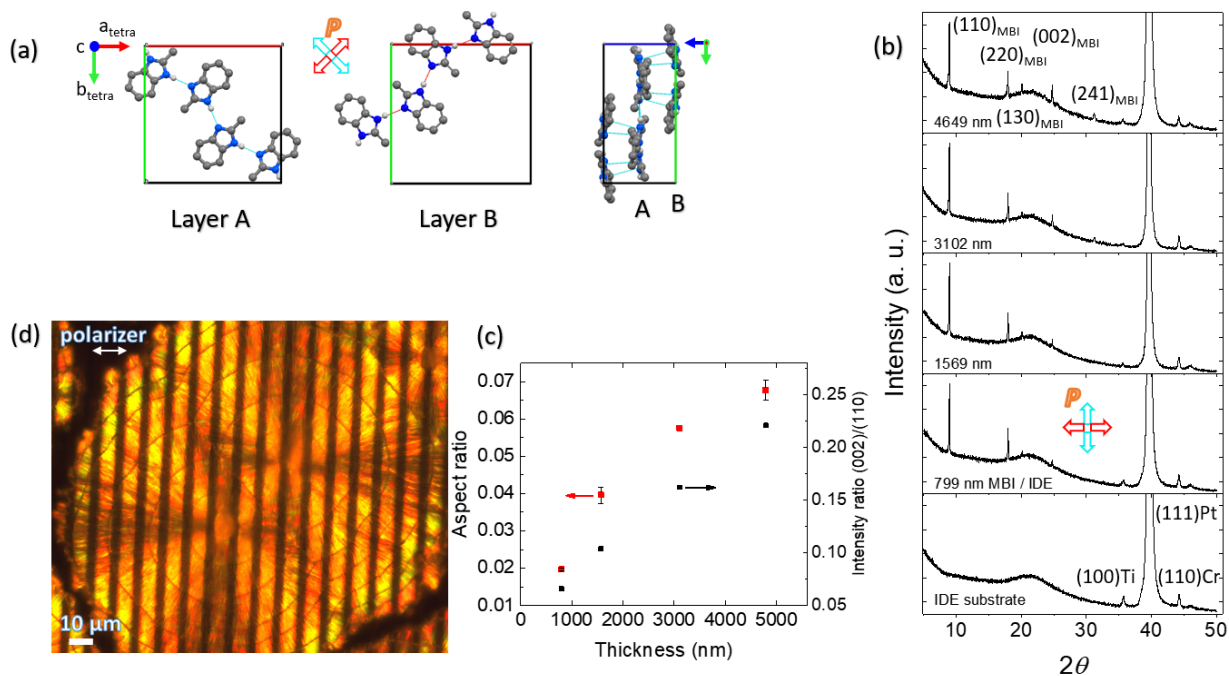

Fig. 4 a) Schematic illustration of molecular packing and hydrogen bonding in an MBI crystal lattice. Principal polarization axes (hydrogen bonds) run parallel to either $[110]_{tetra}$ or the crystallographically equivalent $[1\bar{1}0]_{tetra}$ direction. Light blue dash lines represent the short contact with the interatomic distance lower than the sum of Van der Waals radii. b) XRD pattern of the MBI film with out-of-plane diffraction vector for different film thicknesses. Small Ti and Cr peaks are from the adhesion layers between Pt and glass. c) The aspect ratio and peak intensity ratio as a function of MBI film thickness. The corresponding optical image for different thicknesses is shown in Fig. S12. d) Optical image with crossed polarizers.

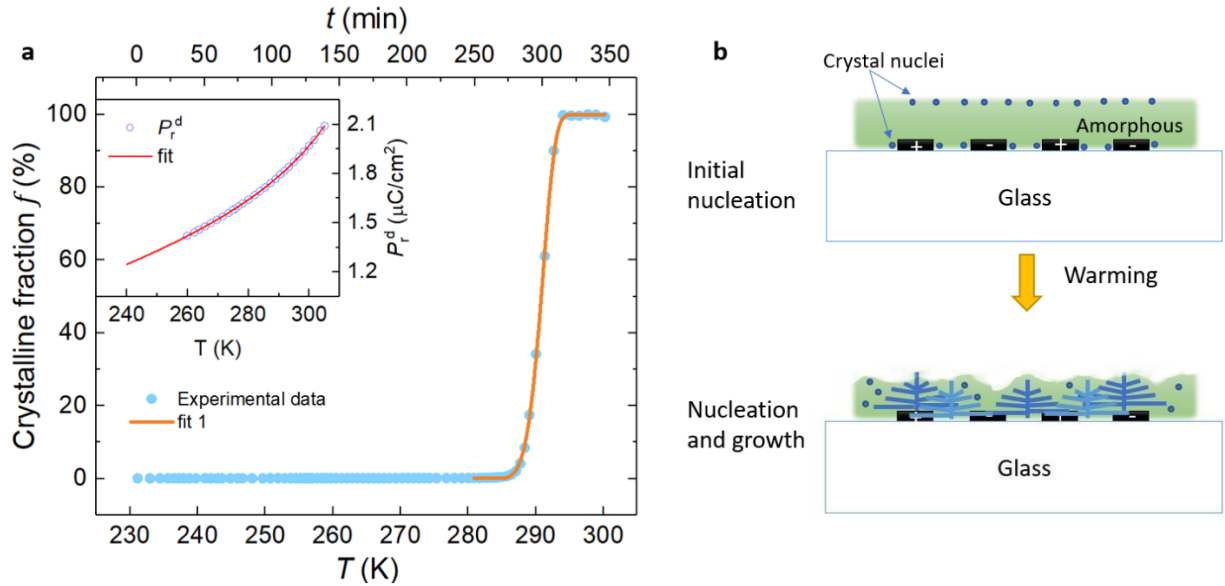

Fig. 5 a) Variation of the crystalline fraction with temperature during annealing. The inset is the variation of remanent polarization with temperature after finishing of phase transition, corresponding to the part in decreasing temperature in Fig. 2(a). The definition of $P_r^d$ and the fit is described in Supplementary Note 6. b) Schematics of the crystal nucleation and growth from amorphous state.

Table 1 Fitting parameters of Eq. (4) used to describe $f(T)$.

|  | $E$ (eV) | $n$ | $T_0$ (K) | $k\tau_0$ (K) |
|---|---|---|---|---|
| Fit 1 | 0.21 | 4.21 | 283.4 | 0.0021 |

Table 2 Properties of some amorphous materials

| Materials | Crystallization temperature $T_x$ (K) | Activation energy in the amorphous-to-crystalline phase transition (eV) |
|---|---|---|
| TaSi$_2$ thin films[26] | 573 | 1.85 |
| Ge$_2$Sb$_2$Te$_5$ films[24] | 398 | 3.2 |
| Si films[14] | 873 | 3.8 |
| Rubrene[20] (C$_{42}$H$_{28}$) | 323 | 0.78 |
| MBI | 287 | 0.21 |

*Supplementary Information*

# Textured organic ferroelectric films from physical vapor deposition and amorphous-to-crystalline transition

## Supplementary Note 1. Electrical measurements for the substrate

In CA/IDE, the contribution from the resistance of substrate glass to the small leakage current is negligible since it is as large as 67 GΩ at 273 K (Fig. S2b).

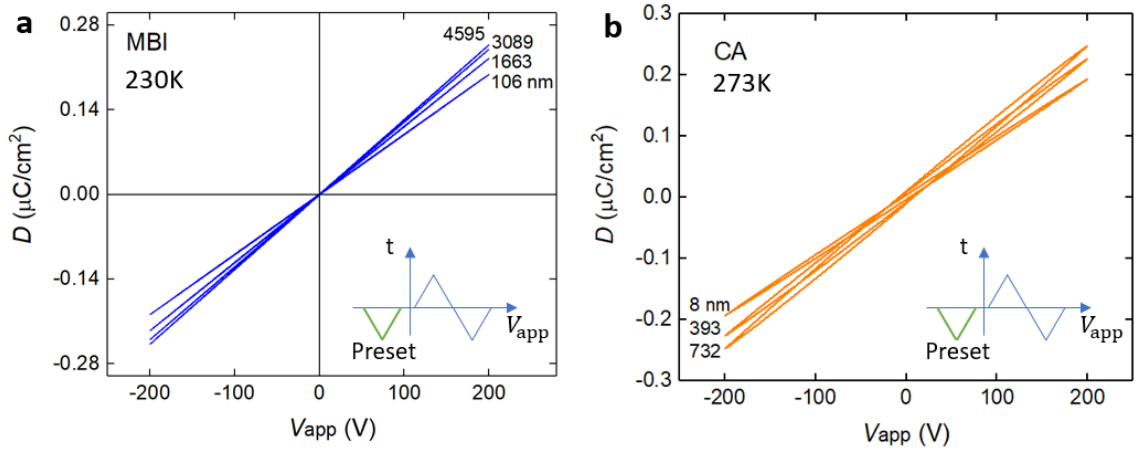

Fig. S1 The D-$V_{app}$ hysteresis loops for (a) MBI measured at 230 K, (b) CA measured at 273 K. All loops were measured in situ at constant temperature in high vacuum (<$1 \times 10^{-6}$ Torr) during deposition process at different thickness.

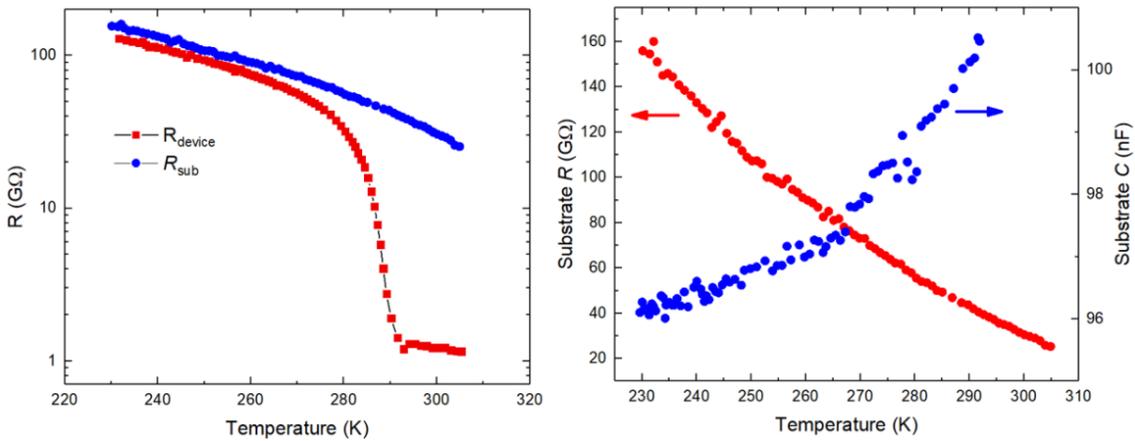

Fig. S2 The variations of resistance of the device and the substrate measured at 200 V with the increasing temperatures. The capacitance of the substrate was measured before the deposition.

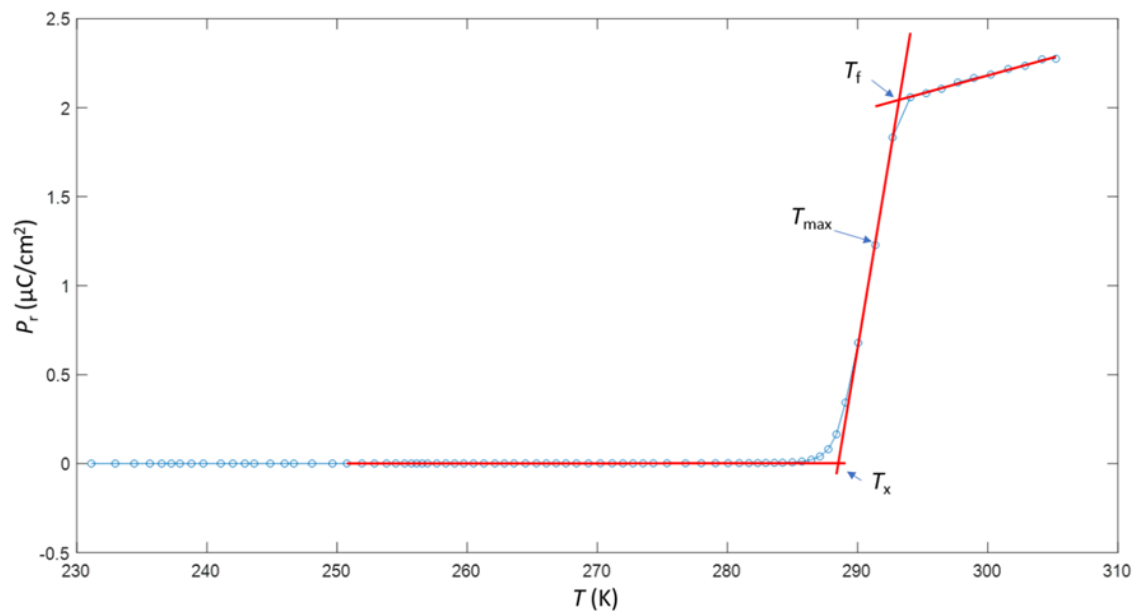

Fig. S3 The crystallization temperature $T_x$ is defined as the temperature at the intersection of the red regression lines which are the linear sections below and above $T_x$. $T_f$ is the end temperature of the crystallization. $T_{max}$ is the temperature with the largest transition rate.

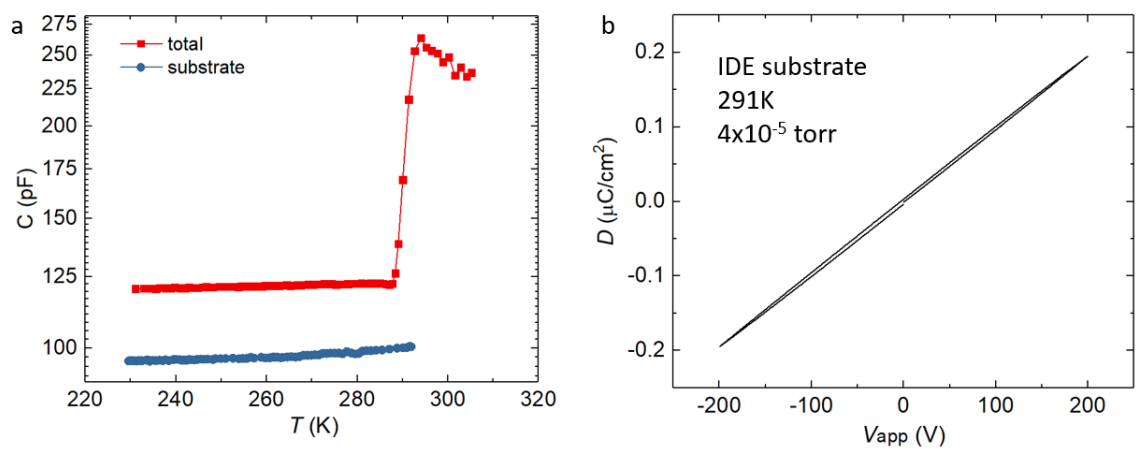

Fig. S4 a) The capacitance of the whole device, and the substrate as a function of temperature. b) the $D$-$V_{app}$ relation of the substrate at 291 K in the vacuum.

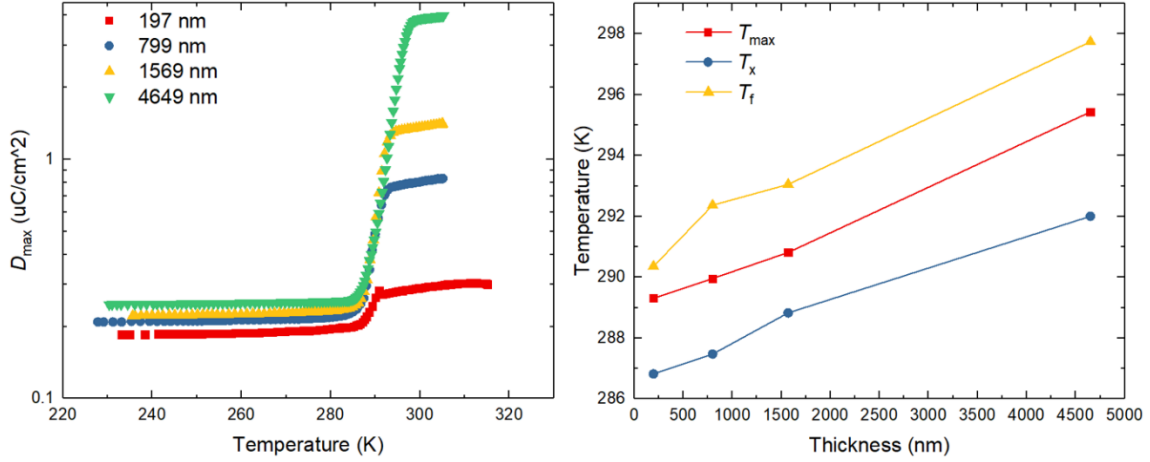

*Fig. S5 Variations of crystallization temperature with the thickness. $T_x$ and $T_f$ are the intersection points of the linear parts of the $D_{max}$ – $T$ curves. $T_{max}$ is the temperature with the largest transition rate.*

## Supplementary Note 2. Calculation of capacitance from non-switching polarization in double-wave method measurements.

In terms of a lossy capacitor for which leakage resistance is in parallel with the capacitance, when a triangle waveform voltage is applied, $V_{app} = k_V t + V_0$, the relationship between the electric displacement and applied voltage can be described by the following equation,

$$D = \frac{V_{app}^2}{2RAk_V} + \frac{C}{A}V_{app} - \frac{V_0^2}{2RAk_V} - \frac{C}{A}V_0 + D_0 \tag{1}$$

where $C$ is the capacitance of the lossy capacitor, $R$ the resistance, $D_0$ the initial displacement, $A$ the electrode area. By fitting Eq. (1) with the non-switching part of the data in Fig. S6, $C_\uparrow$ and $C_\downarrow$ can be obtained, and the final capacitance is the average of $C_\uparrow$ and $C_\downarrow$. We verified that the fitted mean $C$ has the similar value with the capacitance measured by the impedance analyzer (Solartron 1260) shown in Fig. S6.

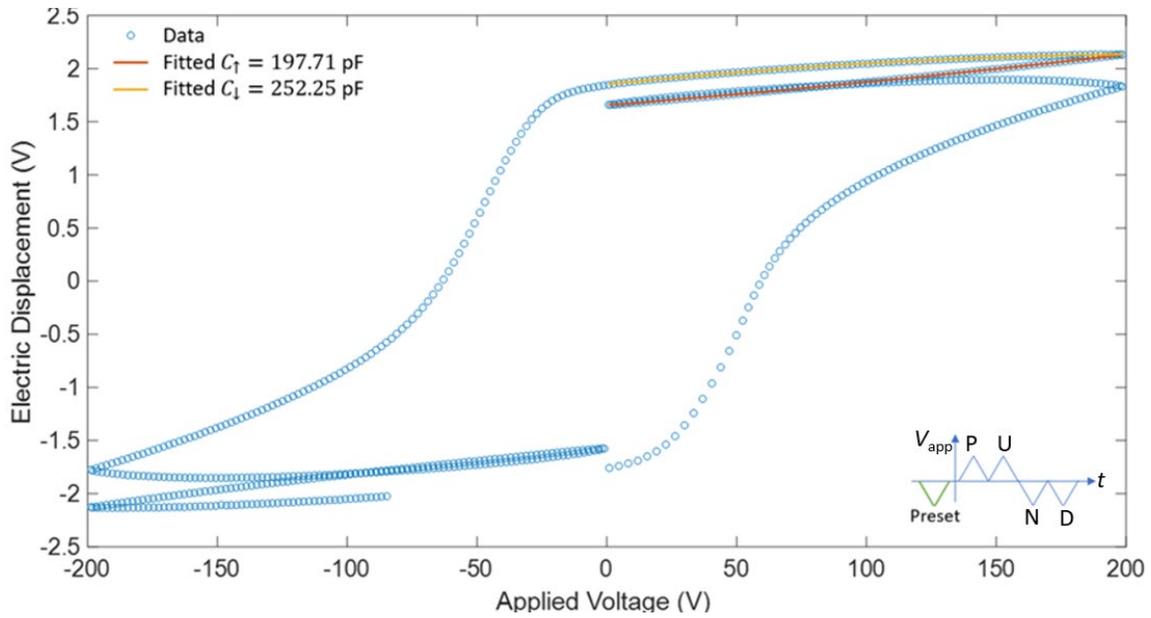

*Fig. S6 The hysteresis loop measured by DWM method at 1 Hz. The mean capacitance is 224.98 pF which is close to the capacitance measured by impedance analyzer (227.33 pF). P: positive, U: up, N: negative, D: down.*

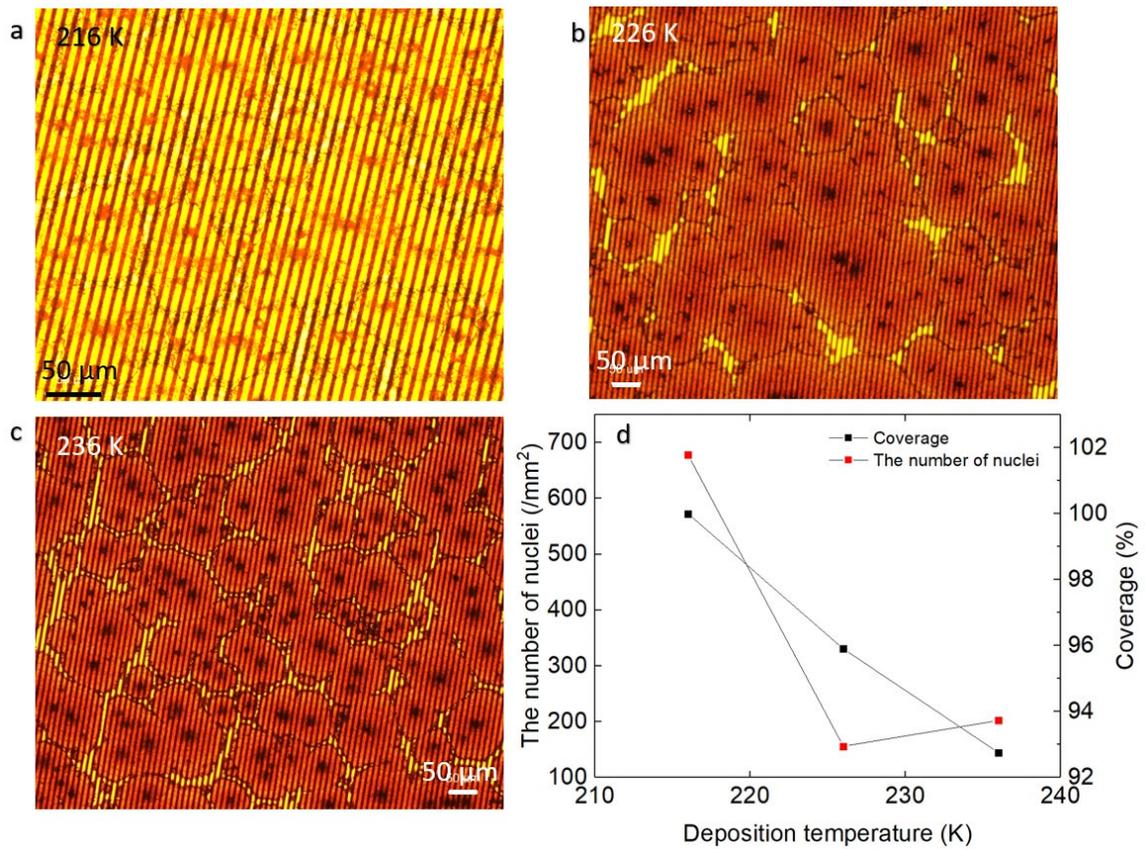

*Fig. S7 a, b, c) The optical image of MBI with the nominal thickness of 1 μm depositted at different temperatures. d) The corresponding nuclei density and coverage as a function of the deposition temperature.*

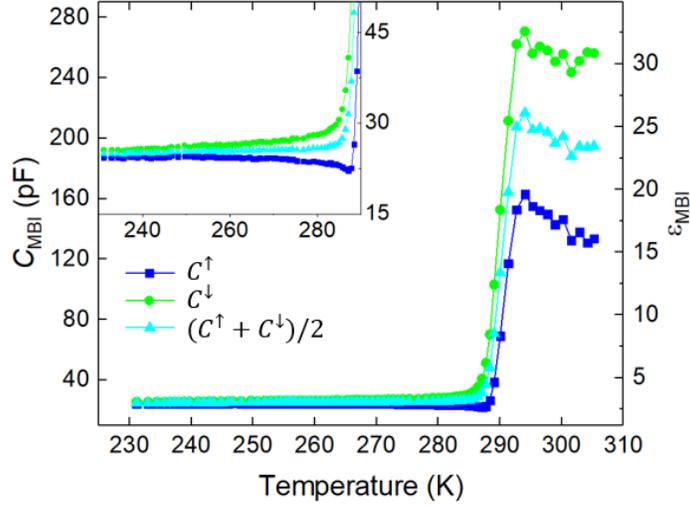

*Fig. S8 Variations of the fitted capacitance with the increasing temperature.*

## Supplementary Note 3, fitted capacitance from the nonswitching part of the DWM loops.

In Fig. S8, $C^\uparrow$ and $C^\downarrow$ are acquired from the fitting of sweep-up and sweep-down of the polarization Up branch respectively (Fig. S6). Although the fitted $D(V)$ functions agree well with the experimental data, the resultant capacitances are much different. Actually, the depolarized field from the remnant polarization established by the Positive branch will suppress the subsequent dielectric polarization by sweeping up, and similarly it will accelerate the dielectric depolarization by sweeping down, so it can be seen that $(C^\downarrow - C^\uparrow)$ increases with the increase in $P_r$ (Fig. S9a). We verified that the average value $(C^\uparrow + C^\downarrow)/2$ is consistent with the capacitance measured by the Impedance Analyzer (Fig. S6) which reduces the impact of $P_r$ by using alternating small voltages.

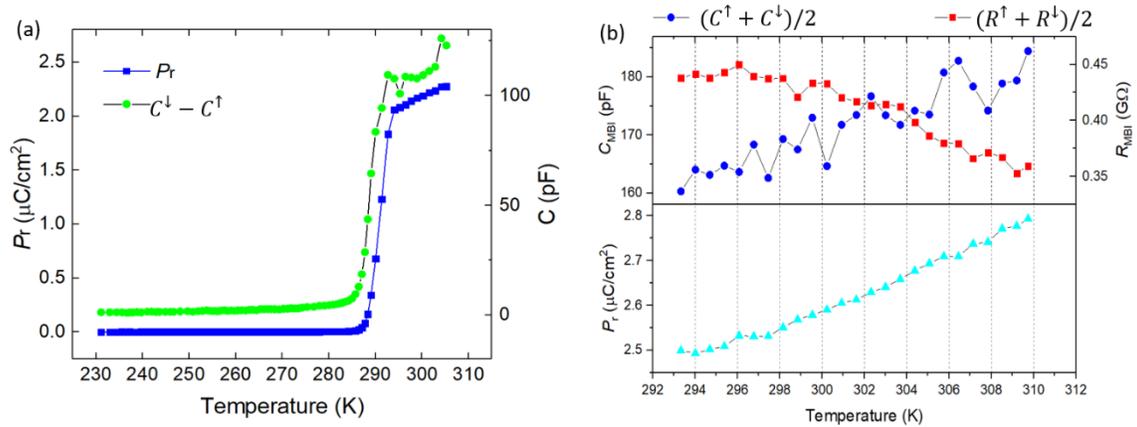

*Fig. S9 a) Variations of remanent polarization with the temperature. b) Variations of capacitance and resistance of the MBI film with the temperature. Both $C_{MBI}$ and $R_{MBI}$ are obtained by fitting the nonswitching part of DWM loops.*

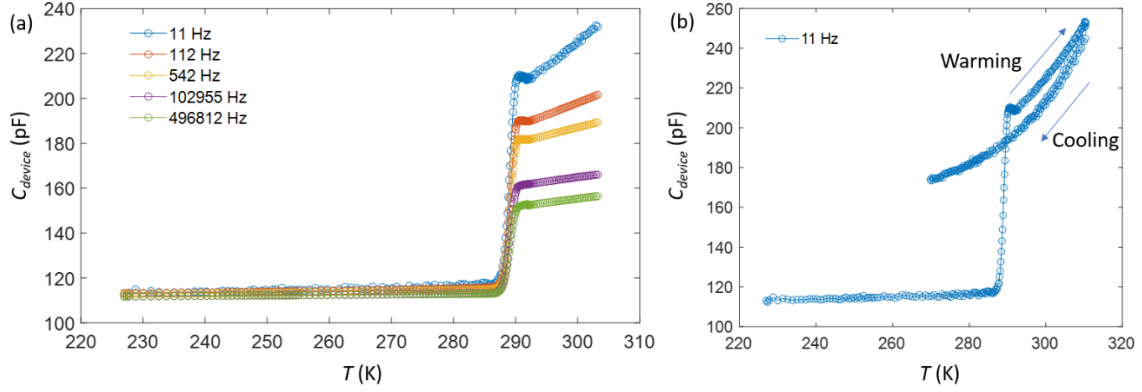

*Fig. S10 a) Variations of capacitance measured by the Solartron 1260 Impedance Analyzer at different frequencies with the temperature. b) Variations of capacitance with the temperature. After the cooling down, the second warming curve overlaps with the cooling curve. The film thickness is 1.55 μm.*

## Supplementary Note 4, the analytical model of the IDE sensor.

The capacitance of the interdigital capacitor is given by[1]:

$$C_{\text{IDC}} = (N-3)\frac{C_{\text{I,IDC}}}{2} + 2\frac{C_{\text{I,IDC}}C_{\text{E,IDC}}}{C_{\text{I,IDC}} + C_{\text{E,IDC}}} \tag{6}$$

$$C_{\text{I,IDC}} = \varepsilon_0 L \left( \frac{K(k_{\text{I}\infty})}{K(k'_{\text{I}\infty})} + (\varepsilon_1 - 1)\frac{K(k_{\text{I},1})}{K(k'_{\text{I},1})} + \varepsilon_S \frac{K(k_{\text{I}\infty})}{K(k'_{\text{I}\infty})} \right) \tag{7}$$

$$C_{\text{E,IDC}} = \varepsilon_0 L \left( \frac{K(k_{\text{E}\infty})}{K(k'_{\text{E}\infty})} + (\varepsilon_1 - 1)\frac{K(k_{\text{E},1})}{K(k'_{\text{E},1})} + \varepsilon_S \frac{K(k_{\text{E}\infty})}{K(k'_{\text{E}\infty})} \right) \tag{8}$$

where $N$ is the number of electrodes, $L$ is the length of the electrode fingers (Fig. S11), $\varepsilon_0$ is the vacuum permittivity, $\varepsilon_1$ and $\varepsilon_S$ are the dielectric constant of film and substrate respectively, $K(k)$ is the complete elliptic integral of first kind with modulus $k$, $k'$ is the complementary modulus through the equation $k' = \sqrt{1-k^2}$, and

$$k_{\text{I}\infty} = \sin\left(\frac{\pi}{2}\eta\right) \tag{9}$$

$$k_{\text{I},1} = \text{sn}(K(k)\eta, k)\sqrt{\frac{\left(\frac{1}{k}\right)^2 - 1}{\left(\frac{1}{k}\right)^2 - [\text{sn}(K(k)\eta, k)]^2}} \tag{10}$$

$$k_{\text{E}\infty} = \frac{2\sqrt{\eta}}{1+\eta} \tag{11}$$

$$k_{E,1} = \frac{1}{\cos h\left(\frac{\pi(1-\eta)}{8r}\right)} \sqrt{\frac{\left[\cos h\left(\frac{\pi(\eta+1)}{8r}\right)\right]^2 - \left[\cos h\left(\frac{\pi(1-\eta)}{8r}\right)\right]^2}{\left[\cos h\left(\frac{\pi(\eta+1)}{8r}\right)\right]^2 - 1}} \quad (12)$$

$$k = \left(\frac{v_2(0, \exp(-4\pi r))}{v_3(0, \exp(-4\pi r))}\right)^2 \quad (13)$$

where $sn(K(k)\eta, k)$ is the Jacobi elliptic function of modulus $k$, $v_2$ and $v_3$ are the Jacobi theta functions, $\eta$ is the metallization ratio through equation $\eta = W/(W+G)$, W is the finger width, G is the gap (Fig. S11), $r = h/(2W+2G)$, $h$ is the thickness of the film. In summary, the capacitance of IDE sensor is a function of $\eta$, $r$, $\varepsilon_1$ and $L$ (i.e., $C_{IDC} = C_{IDC}(\varepsilon_1, L, \eta, r)$).

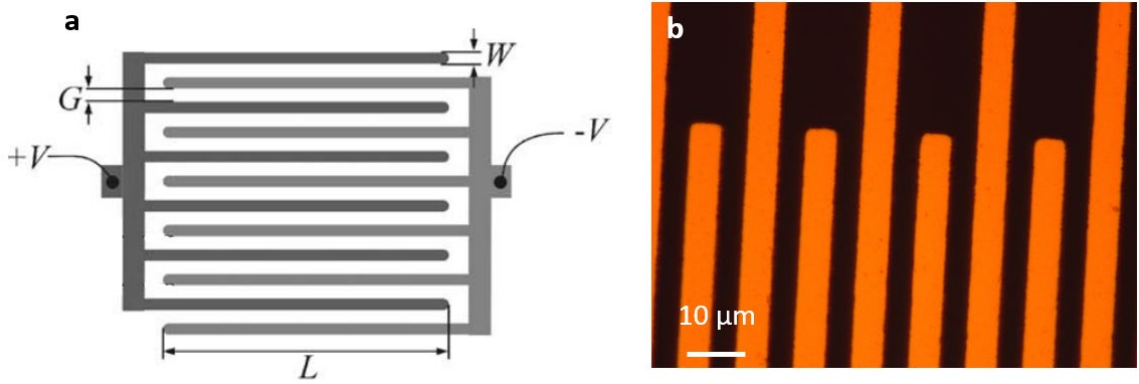

Fig. S11 (a) The dimensions of the interdigitated electrodes.[1] (b) The optical image of the substrate.

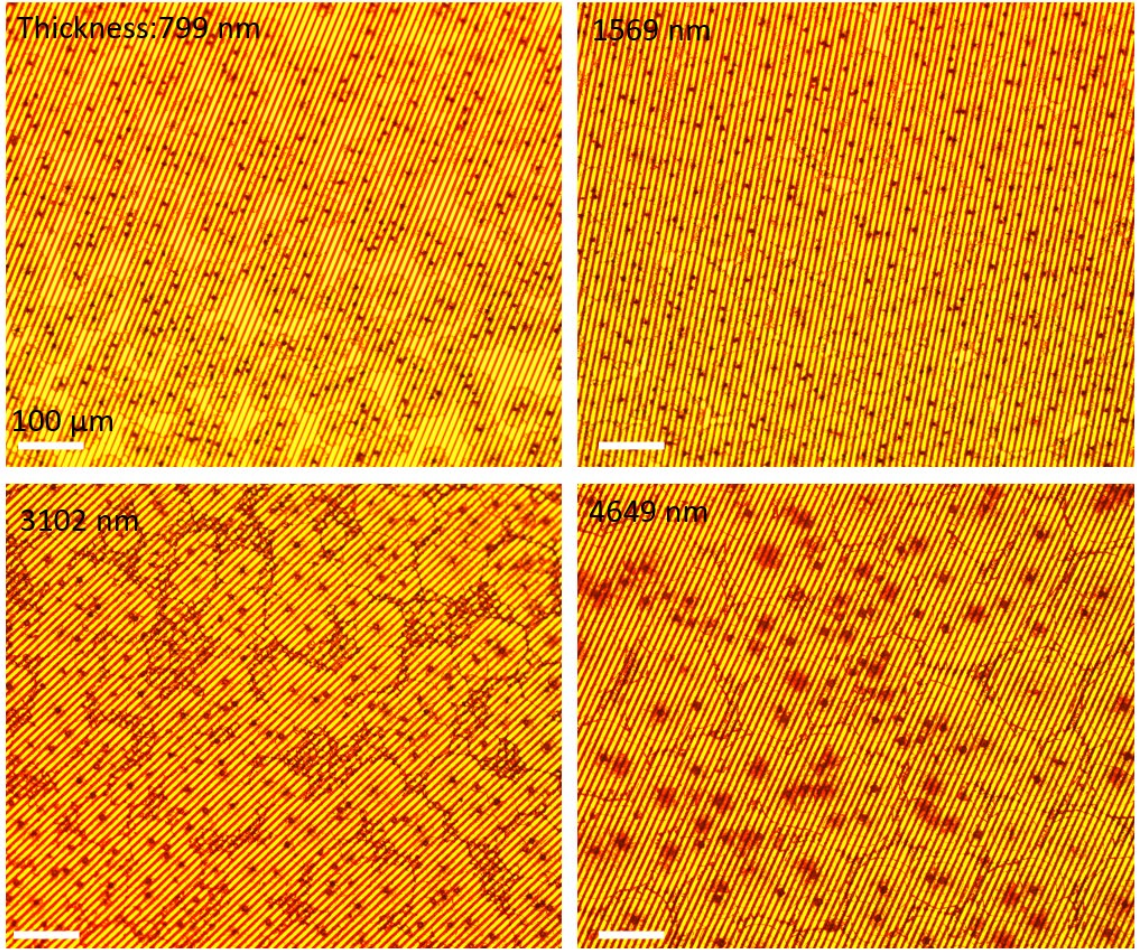

*Fig. S12 The optical image of MBI films on IDE substrate, deposited at 230 K with different thicknesses. All scale bara are 100 μm. The black dots are counted to calculate nuclei density.*

## Supplementary Note 5. Reflectance observed using polarizer and analyzer.

Angle $\alpha$ between the optical c axis ($\hat{x}$ axis) and the polarizer axis. The transmittance along a and c axis is $t_a$ and $t_c$.

The incident light passing through the polarizer has the polarization as

$$\vec{E} = E(\cos\alpha\,\hat{x} + \sin\alpha\,\hat{y})$$

After passing through the crystal, the polarization becomes

$$\vec{E}_1 = E(t_c \cos\alpha\,\hat{x} + t_a \sin\alpha\,\hat{y})$$

The analyzer axis, which is perpendicular to the polarizer axis, is

$$\hat{A} = (\sin\alpha\,\hat{x} - \cos\alpha\,\hat{y}).$$

After the analyzer the light amplitude becomes:

$$\vec{E}_1 \cdot \hat{A} = \frac{1}{2}E(t_c - t_a)\sin(2\alpha).$$

When $\alpha = n\left(\frac{\pi}{2}\right)$, there is no transmitted light.

The maximum light is expected when $2\alpha = \frac{\pi}{2} + n\pi$ or $\alpha = \frac{\pi}{4} + \frac{n\pi}{2}$.

## Supplementary Note 6

As shown in the Fig. 5a inset, we firstly fit the relationship $P_r^d(T)$ with a custom function, $P_r^d(T) = 0.2763 \exp(0.006261 * T) + 7.315 * 10^{-10} \exp(0.06409 * T)$. All remanent polarization values are normalized to those at 300 K through the equation, $P_r^{'}(T) = P_r(T) * P_r^d(300)/P_r^d(T)$. The crystalline fraction is expressed as, $f = P_r^{'}/P_{r0}^{'}$, where $P_{r0}^{'}$ is the maximum value among $P_r^{'}(T)$.